\documentclass[]{spie}  

\usepackage{amsmath,amsfonts,amssymb}
\usepackage{graphicx}
\usepackage{subfigure}
\usepackage[colorlinks=true, allcolors=blue]{hyperref}
\usepackage{svg}
\usepackage{gensymb}
\usepackage{amssymb}
\usepackage{wasysym}
\usepackage{textcomp}
\usepackage{tcolorbox}

\title{A multi-object approach for studying exoplanet atmospheres using high-resolution spectrographs}

\author[a,b]{Manjunath Bestha}
\author[c]{Athira Unni}
\author[a]{T. Sivarani}
\author[a]{Dhanush S R}
\author[a]{Lokesh Manickavasaham}
\author[a,b]{Parvathy M}
\author[d]{Devika K Divakar}
\author[a]{Arun Surya}

\affil[a]{Indian Institute of Astrophysics, Bangalore, India}
\affil[b]{University of Calcutta, India}
\affil[c]{University of California, Santa Cruz, USA}
\affil[d]{University of Texas, Austin, USA}

\authorinfo{Further author information(Send correspondence to Manjunath Bestha): \\ Manjunath Bestha: E-mail: bestha95@gmail.com\\}

\begin{document} 

\maketitle

\begin{abstract}

Atmospheric characterization of exoplanets has traditionally relied on Low-Resolution Transmission Spectroscopy (\textsc{Lrts}), obtained from both space- and ground-based facilities, as well as on High-Resolution Transmission Spectroscopy (\textsc{Hrts}). Although \textsc{Hrts} can resolve individual spectral lines, it is subject to normalization degeneracies that limit the accurate retrieval of key atmospheric parameters such as pressure, abundance, and cloud opacity. A promising strategy to mitigate this issue is to combine ground-based \textsc{Hrts} with space-based \textsc{Lrts}. However, this approach depends on two separate datasets, thereby requiring two independent observations. In this study, we explore the feasibility of Multi-Object High-Resolution Transmission Spectroscopy (\textsc{Mo-Hrts}) as a means to constrain atmospheric parameters in retrievals using a single dataset. Through simulations based on existing spectrograph specifications for a well-studied target, we demonstrate that low-resolution broadband transmission spectra can be extracted from \textsc{Mo-Hrts} data.

\end{abstract}


\section{INTRODUCTION}
\label{sec:intro}  

Exoplanet atmospheric characterization relies on both space-based transmission spectroscopy and ground-based Low-Resolution and High-Resolution Transmission Spectroscopy (\textsc{Lrts}  \& \textsc{Hrts}). \textsc{Hrts}  can resolve the position and shape of individual spectral lines, which can provide information about atmospheric circulation, temperature-pressure profile, and atmospheric parameters like the equilibrium temperature, metallicity, and abundances of different elements \cite{Komacek_2016,João, Wakeford_2017} while  \textsc{Lrts} is effective for detecting broad absorption features from both atomic and molecular species and is particularly crucial for detecting clouds and hazes, which preserves the spectral continuum. 

However, ground-based \textsc{Lrts}  and \textsc{Hrts}  observations often suffer from the Earth's atmospheric and instrumental systematics, which introduce ambiguities in detected signals and affect the precision of measurements\cite{Jiang}. To mitigate these issues, ground-based \textsc{Lrts}  uses a comparison star to remove systematics via common-mode corrections. Additionally, Gaussian Process Regression (\textsc{Gpr}) can model wavelength-dependent systematics, leading to a more precise measurement of transit parameters such as the planet-to-star radius ratio ($R_p/R_*$) in spectroscopic light curves. Although \textsc{Gpr}  can be effective without a comparison star, incorporating one has been shown to increase the precision of measurements \cite{Vatsal}. 

In \textsc{Hrts}, the analysis involves multiple normalization steps—first, the continuum normalization of each echelle order, and second, the division of in-transit spectra by out-of-transit spectra \cite{Wyttenbach}. Additionally, instrumental systematics such as spectral “wiggles” are often corrected using spline-based normalization techniques, as demonstrated in \textsc{Espresso}  data \cite{Tabernero}. These steps introduce a normalization degeneracy in the retrieval of spectral lines, wherein the inferred molecular abundances and cloud properties become correlated. As a result, the shapes of spectral lines alone do not provide sufficient information to fully constrain atmospheric parameters such as temperature, pressure, and composition\cite{Fisher_2020}. These normalization-induced degeneracies limit the study of atmospheric escape by introducing significant uncertainties in the pressure, temperature, and origin of upper-atmosphere absorption features. This, in turn, complicates the modeling of planetary evolution and the robust estimation of mass loss rates \cite{Linssen}. To overcome the challenges in \textsc{Hrts}, a combined approach using space-based \textsc{Lrts}  and ground-based \textsc{Hrts}  has been used \cite{Brogi_2017}. However, this method is based on two separate data sets \cite{Gloria,asi_abs}, thereby necessitating two independent observations. In this work, we are proposing a promising observation strategy to simultaneously perform an \textsc{Lrts} and an \textsc{Hrts}.

The proposed approach constructs a low-resolution broadband transmission spectrum (wavelength vs.\ $R_p/R_*$, as shown in Figure~\ref{fig:lrts}) directly from high-resolution data, utilizing spectrophotometric techniques commonly employed in ground-based, multi-object low-resolution transmission spectroscopic (\textsc{Mo-Lrts}) observations. 
Constructing the broadband transmission from high-resolution data has previously been demonstrated by Snellen~\cite{snellen2004new}, who utilized the chromatic Rossiter--McLaughlin effect with highly stable spectrographs. 
In addition, we present an approach to constructing the broadband transmission spectra using spectrographs with moderate stability. The method utilizes the simultaneous observation of both the target exoplanet host star and a comparison star. To our knowledge, the utilization of a multi-object, broadband high-resolution spectrograph for transmission spectroscopy has not been reported in the literature.

Among current instruments, only a few—such as \textsc{Uves/Flames} and \textsc{Giraffe/Flames} at the Very Large Telescope (\textsc{VLT}), and \textsc{Ghost}  on Gemini—are capable of simultaneous multi-object high-resolution spectroscopy with broad wavelength coverage suitable for our approach. One among them is the Fibre Large Array Multi Element Spectrograph (\textsc{Flames}), mounted on \textsc{VLT}, which offers two modes: \textsc{Uves/Flames} (Ultraviolet and Visual Echelle Spectrograph) and \textsc{Giraffe/Flames} Spectrograph. \textsc{Uves/Flames} supports simultaneous observations of up to eight targets, while \textsc{Giraffe/Flames} allows observations of up to 130 targets. Within \textsc{Flames}, both the \textsc{Giraffe} and \textsc{Uves}  spectrographs have previously been used for transmission studies of exoplanets. However, \textsc{Giraffe} is less optimal for our application, since it does not provide broad wavelength coverage in a single exposure; multiple setup observations are needed to obtain broadband data, which makes constructing broad transmission spectra more challenging. Alongside this, \textsc{Uves/Flames} is currently one of the very few operational multi-object high-resolution spectrographs capable of simultaneous observations of multiple targets along with the program target. It provides a spectral coverage of approximately 200 nm at a resolution of \( R \approx 47{,}000 \)\footnote{\url{https://www.eso.org/sci/facilities/paranal/instruments/flames/overview.html}} for all targets, allowing us to detect the expected orbital velocity trend of the planet in the stellar rest frame. This capability makes it suited for implementing and validating the proposed methodology.

Given the proven suitability of \textsc{Uves}  for transmission spectroscopy and its capability for simultaneous observation of the program and multiple comparison stars, we have chosen this instrument as the basis for our study. Our simulations considered specifications of \textsc{Uves/Flames} to demonstrate the feasibility of constructing broadband \textsc{Lrts}, derived from high-resolution data, and the advantages via simultaneous multi-object spectroscopy.

\section{Simulations}

We simulated the high-resolution data using the \textsc{Uves/Flames} specifications and constructed both narrow- and broadband transmission spectra to test the feasibility of this method for the well-studied target \textsc{wasp}-121 b. The approach is intended to be extended to other targets as well as to additional spectrographs on different facilities, such as the Multi-object Spectroscopic Mode (\textsc{MoS}) of the High-Resolution Optical Spectrograph (\textsc{Hros}) on the Thirty Meter Telescope (\textsc{TMT} )~\cite{Manju_ADC,Manjunath,devika_akka}.

We have selected \textsc{wasp}-121 b based on several key factors. First, it is an exoplanet with a well-characterized atmosphere, enabling comparison of our results with existing studies\cite{Jaime_W,Merritt,Langeveld}. Second, the planet’s orbital velocity is sufficiently high to be resolved by the \textsc{Flames} spectrograph. Finally, suitable bright comparison stars are present within the \textsc{Flames} field of view, facilitating differential analyses.

\subsection{Data Simulation}

\textsc{wasp}-121 b is an ultra-hot Jupiter with a radius of 1.865 $R_{Jup}$ in a 1.2749-day orbit around an F6-type main-sequence star of V magnitude 10.52. Radial-velocity follow-up confirmed the planet, with its mass measured at 1.184 $M_{Jup}$ and an equilibrium temperature of about 2350 K. The large transit depth (12.355 ppt), the stellar rotational velocity ($v\sin i = 13.5 \pm 0.7$ km s$^{-1}$ \cite{Bourrier_2020}), and the relatively high Transmission Spectroscopy Metric (TSM = 367.2)\footnote{\url{https://research.iac.es/proyecto/exoatmospheres/table.php}} underline its suitability as an atmospheric target. Although the TSM is primarily defined for space-based observations, it still provides a useful indicator of atmospheric observability. Together with the planet’s high orbital velocity of 218 km s$^{-1}$ around the host star \cite{Stephanie}, these properties make \textsc{wasp}-121 b particularly favorable for our method. The planet has also been studied with both high- and low-resolution ground-based instruments, such as \textsc{Harps}-N \cite{Seidel_2023} and \textsc{Gmos} on Gemini \cite{Jaime_W}, which will provide valuable baselines for comparison with our measurements. 

\begin{figure}[h!]
    \centering
    \includegraphics[width=1\textwidth]{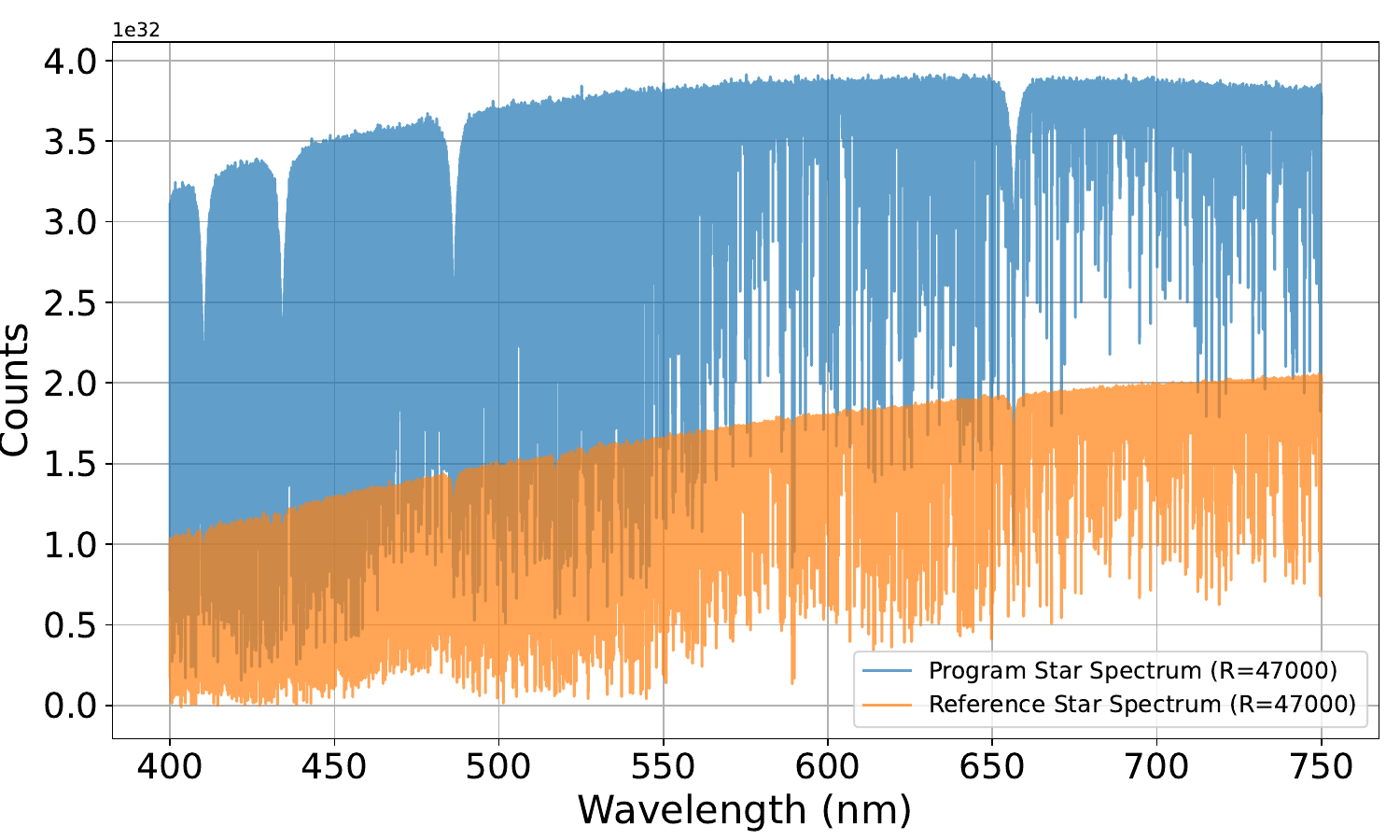}
    \caption{Simulated high-resolution spectra of the program star \textsc{wasp}-121 (an F-type star) and a G-type comparison star with a similar magnitude. The spectral resolution is degraded to 47,000, consistent with the \textsc{Flames} spectrograph.}
    \label{fig:spectra}
\end{figure}

\begin{figure}[h!]
    \centering
    \includegraphics[width=1\textwidth]{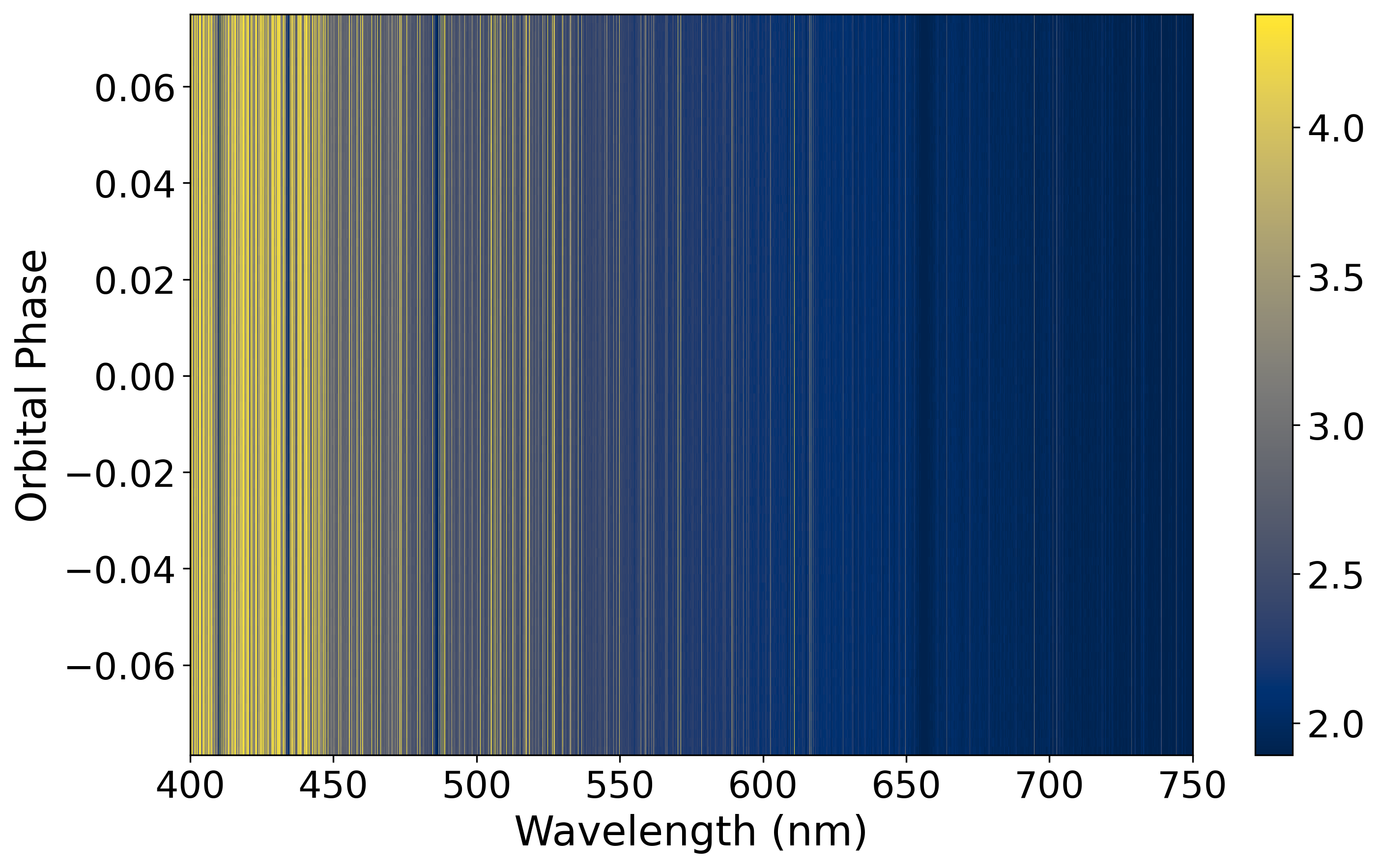}
    \caption{The image shows the data of a simulated planetary signal, and noise has been injected into the residuals of the stellar spectra, corresponding to a specific orbital phase of the planet.}
    \label{fig:sim_data}
\end{figure}

In our simulation, we modeled the spectra of both the program (F-type) and comparison stars observed by Jaime et al. \cite{Jaime_W}. As the coordinates of the comparison star were not reported in their study, we have chosen the nearest bright star (G-type) to the program target within the \textsc{Gmos} field of view and adopted it as the reference for our analysis. The \textsc{Uves/Flames} red arm setup, centered at 580 nm, provides a 200 nm bandwidth. For our simulations, however, we considered a broader wavelength range (400 - 750 nm) to focus on the sodium doublet and the Rayleigh slope, which are frequently detected in ultra-hot Jupiters. These features probe the upper atmospheric layers and are important for studying atmospheric composition, thermal gradients, wind velocities, and escape mechanisms.

We obtained high-resolution stellar spectra using the \texttt{expecto}\footnote{\url{https://pypi.org/project/expecto/}} library. 
For the program star, the spectrum was generated assuming an effective temperature of 
$T_{\mathrm{eff}} = 6776$~K and a surface gravity of $\log g = 4.242$. 
For the comparison star, we considered $T_{\mathrm{eff}} = 5600$~K and $\log g = 4.44$. 
The model fluxes were subsequently converted into photon counts using the telescope’s 
collecting area, exposure time, and overall throughput.

To match the instrumental resolution, the modeled spectra were convolved with a Gaussian kernel, where the full width at half maximum (FWHM) was given by
\[
\mathrm{FWHM} = \frac{\lambda}{R},
\]
with $\lambda$ representing the mean wavelength and $R$ the instrumental resolution, as the resolution of the modeled spectrum is constant across the wavelength band.

Photon noise was added to simulate observational conditions, with a target signal-to-noise ratio (SNR) of 100 per resolution element for the program star and 80 for the comparison star. 
The difference in SNR accounts for expected variations in fiber throughput and differential losses between the two targets. 
The noise-added spectra of both stars are shown in Figure~\ref{fig:spectra}. 
By dividing the noise-affected flux of the program star by that of the comparison star, we obtained the residual stellar spectrum.

To include the impact of differential systematics in a \textsc{Mo-Hrts} data set sampled at 82 orbital phases, we generated fiber-specific noise to represent variations in fiber transmission and instrument coupling. For the program star, we added Gaussian noise at the level of $0.5\%$, together with a mild wavelength-dependent modulation of amplitude $\pm 0.02\%$, modeled as a sinusoidal function. For the comparison star, a slightly higher fiber-level noise of $0.8\%$ was introduced, with a similar modulation but represented using a cosine function to mimic a different chromatic response. In reality, fiber transmission variations are more irregular and may not follow exact sinusoidal or cosine patterns; our use of simple functions here is only a convenient way to simulate differential chromatic effects. The ratio of the fiber-specific noises was then used to construct a noise matrix as a function of wavelength and orbital phase.

Airmass effects and telluric absorption features are also not explicitly accounted for, since our analysis is differential in nature. These effects are expected to influence both the program and comparison stars similarly and, therefore, effectively cancel out.  Additionally, our simulation of the stellar spectrum did not include the center-to-limb variation (CLV) or the Rossiter-McLaughlin (RM) effect. While these effects can, in principle, be modeled and corrected \cite{is_na}, they can introduce systematic signals that mimic planetary atmospheric features. Including them requires detailed stellar models and precise system parameters. For this initial proof-of-concept simulation, we chose not to include them so that we could focus on construction of broadband transmission spectra using multi-object high resolution data.  A careful study of the CLV and RM effects will be an important step in future work.

To simulate the planetary transmission spectrum, we employed the \texttt{petitRADTRANS}\cite{Mollière}\footnote{\url{https://petitradtrans.readthedocs.io/en/latest/content/notebooks/retrieval_spectral_model.html}} radiative transfer package. The stellar rest-frame velocity shift and system radial velocity were set to zero, under the assumption that these corrections have already been applied to the simulated data. Considered input values sodium (Na) mass fraction is $\log_{10}(X_{\mathrm{Na}}) \approx -5.7$,  
the opaque cloud top pressure is $\log_{10}(P_{\mathrm{cloud}} \, [\mathrm{bar}]) = -2$,  
and the reference pressure is $\log_{10}(P_{\mathrm{ref}} \, [\mathrm{bar}]) \approx -1.7$
. All other atmospheric and planetary parameters were taken from the default planet model provided by \texttt{petitRADTRANS}. Finally, the simulated data shown in Figure~\ref{fig:sim_data} are the product of the noise matrix, the stellar spectrum after comparison division, and the planetary transmission spectrum at each orbital phase (see Eq.~\ref{sd}).

\begin{equation}
F_{\mathrm{sim}}(\lambda, \phi) = N(\lambda, \phi) \,\times\, 
F_{\star}(\lambda) \times T_{p}(\lambda, \phi),
\label{sd}
\end{equation}

\noindent
\textbf{Where:}
\begin{align*}
F_{\mathrm{sim}}(\lambda, \phi) &\quad \text{: Simulated multi-object data }\\ 
                                &\quad \text{as a function of wavelength } \lambda \text{ and orbital phase } \phi \\
N(\lambda, \phi)                &\quad \text{: Noise matrix} \\
F_{\star}(\lambda)              &\quad \text{: Stellar spectrum after comparison division} \\
T_{p}(\lambda, \phi)            &\quad \text{: Planetary transmission spectrum}
\end{align*}

\subsection{Narrow band transmission spectra}

After generating the simulated high-resolution data, each in-transit spectrum was divided by the mean out-of-transit spectrum, removing features common to both in- and out-of-transit spectra, such as residual stellar signals, and thereby isolating the planet’s atmospheric signature \cite{athira_hires}. Moreover, when the program and comparison stars have similar spectral types and magnitudes, division by the comparison spectrum alone can efficiently reduce stellar contamination and time-correlated systematics common to both stars, ensuring consistency across spectra obtained at different epochs.

To account for the change in radial velocity of the planet during transit, we shifted all in-transit spectra to the planet’s rest frame. The relative velocity offsets for each phase were computed using the \texttt{petitRADTRANS} package, based on the known orbital parameters of the system. This step ensures that the planetary absorption lines are aligned in wavelength space.

\begin{figure}[h!]
    \centering
    \includegraphics[width=1\textwidth]{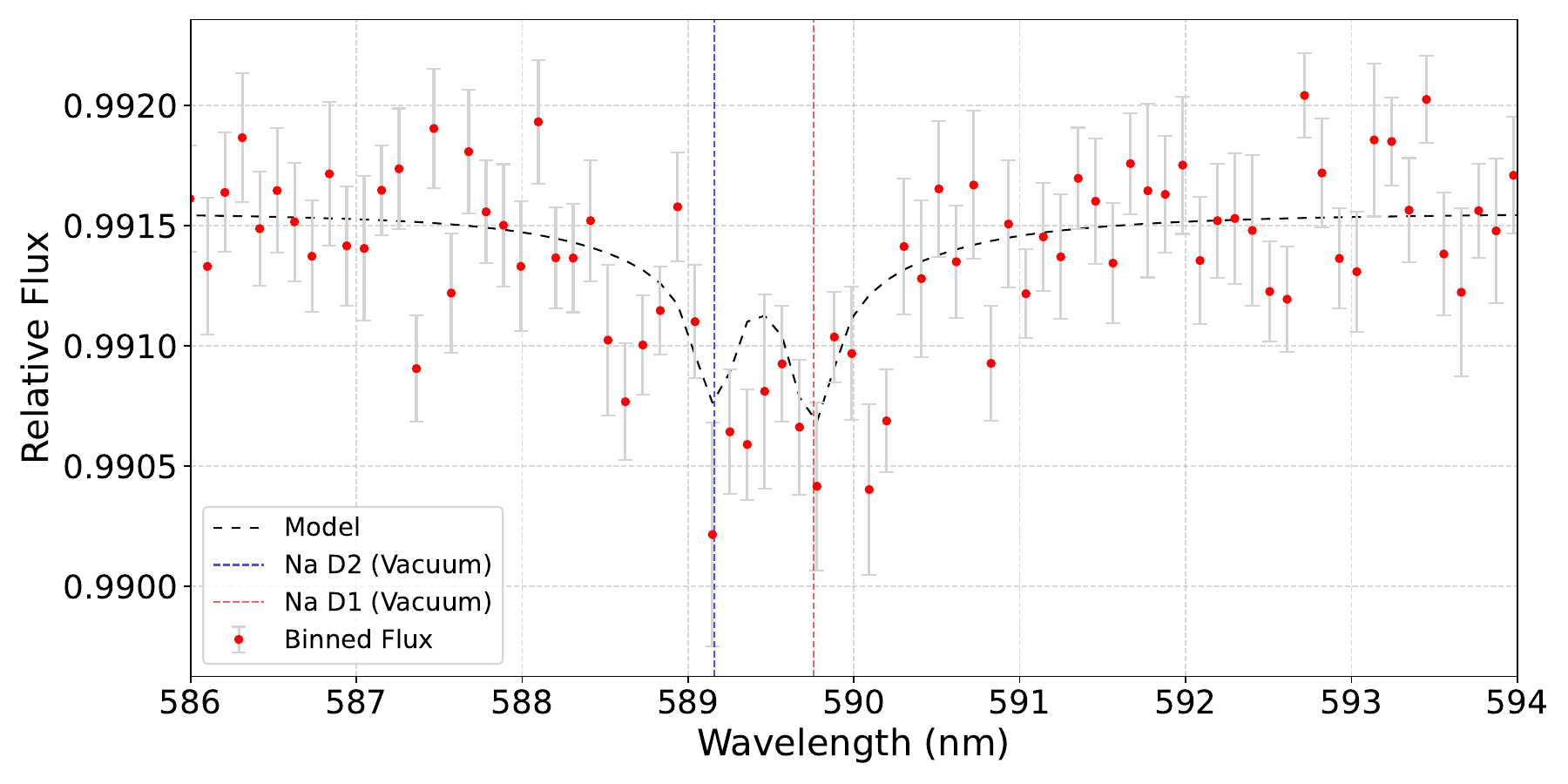}
    \caption{high-resolution transmission spectrum of \textsc{wasp}-121 b. The red points represent the binned spectroscopic data generated from our simulation, which includes realistic noise and systematics. The gray dotted line shows that the noiseless input model spectrum has been shifted to the planet's rest frame and averaged over phase for representation. The blue and red dotted vertical lines mark the positions of the Na D2 and D1 lines, respectively.}
    \label{fig:hrts}
\end{figure}

Once aligned, we computed the average of all shifted in-transit spectra to obtain the high-resolution transmission spectrum (see Figure \ref{fig:hrts}). To further enhance the visibility of spectral features and suppress noise, the transmission spectrum was binned into thirty discrete wavelength points with a bin width of $0.105,\text{\AA}$, clearly showing the Na D1 and D2 absorption features in the simulated planetary transmission spectrum.

\subsection{Broad band transmission spectra}

To construct the low-resolution broadband spectra, we generated a white light curve from the simulated data. This was achieved by summing the flux over the entire spectral range (i.e., integrating along the dispersion direction) and then normalizing the resulting light curve by the median of the out-of-transit flux values. 

In a similar manner, we produced spectroscopic light curves by dividing the spectrum into wavelength bins of 50\,\AA\ and, for each bin, following the same procedure as for the white light curve: summing the flux within the bin and normalizing by the corresponding out-of-transit median. The white light curve was then fitted using the \texttt{Pylightcurve}\footnote{\url{https://pypi.org/project/pylightcurve/}} package to derive the best-fit transit model. From this fit, we obtained the residuals, which represent time-dependent systematics affecting all spectroscopic light curves in a similar way. 

These residuals were subsequently subtracted from each spectroscopic light curve using the \emph{Common Mode Correction} (\textsc{CMC}) technique\cite{unni2024low}. The CMC removes dominant wavelength-independent (achromatic) systematics—typically arising from atmospheric transparency variations, instrumental drifts, or guiding errors—thereby improving the precision of the spectroscopic light curves for transmission spectrum extraction.

\begin{figure}[h!]
    \centering
    \includegraphics[width=1\textwidth]{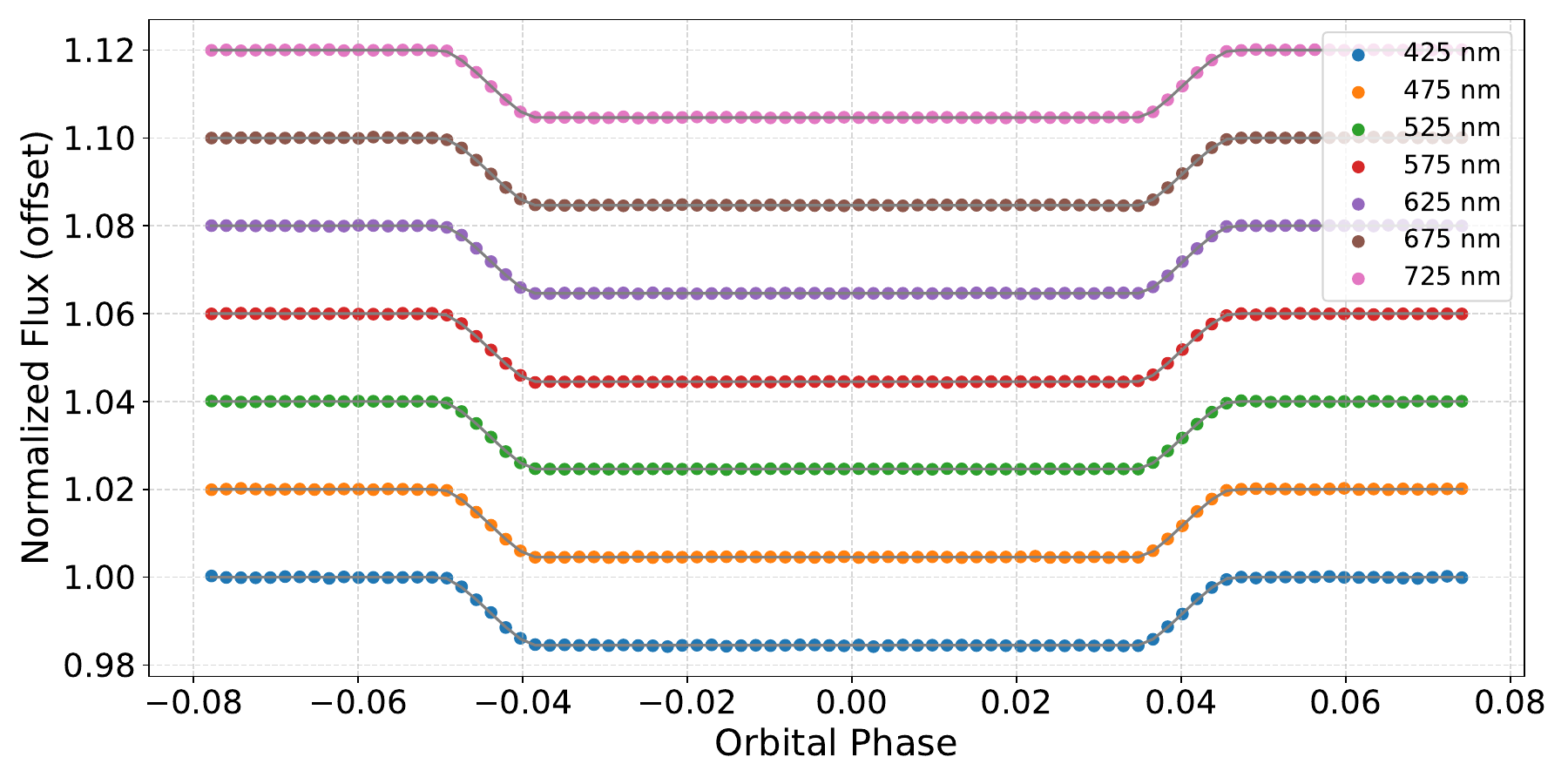}
    \caption{Spectroscopic light curves for \textsc{wasp}-121 b in each wavelength bin (scattered points), with the best-fit transit models overlaid (solid lines). The light curves are vertically offset by 0.01 for clarity. Each bin spans 50\,\AA\ in wavelength.}

    \label{fig:slc}
\end{figure}

\begin{figure}[h!]
    \centering
    \includegraphics[width=1\textwidth]{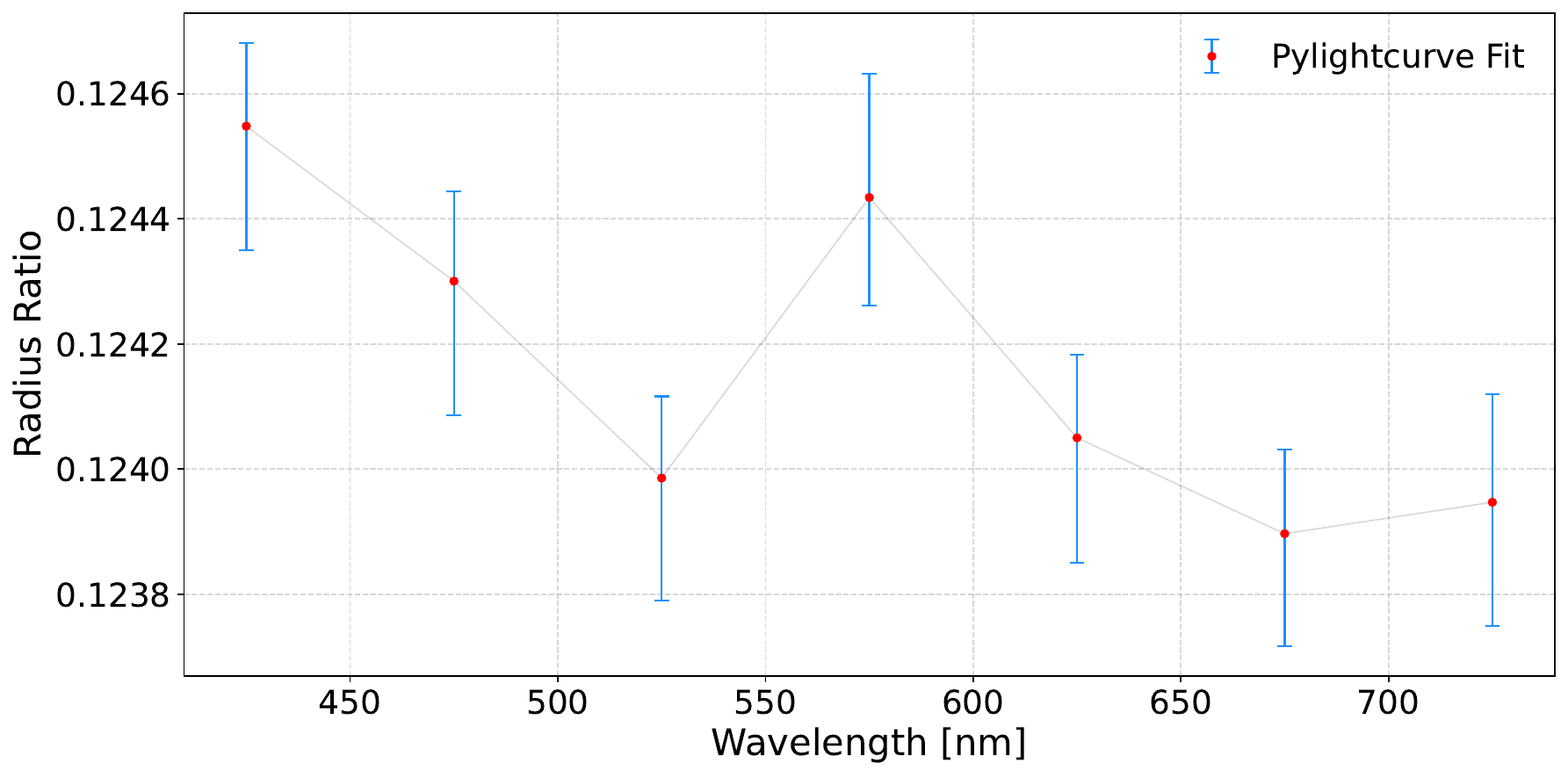}
    \caption{Low-resolution broadband transmission spectrum of \textsc{wasp}-121 b, constructed from high-resolution spectroscopic data. Each scattered red point represents the transit depth estimated using \texttt{PyLightcurve}, while the blue vertical lines indicate the associated uncertainties for each spectroscopic transit light curve.}

    \label{fig:lrts}
\end{figure}

The corrected, wavelength-dependent light curves were fitted using \texttt{pylightcurve} (see Figure \ref{fig:slc}), following the same procedure applied to the white-light curve. The resulting broadband transmission spectrum is shown in Figure \ref{fig:lrts}, which presents the measured transit depths across different spectroscopic light curves. The spectrum appears to show a Rayleigh scattering slope extending up to the bin centered around 525 nm, followed by a pronounced peak in the sodium region, attributed to the increased transit depth caused by the presence of sodium in the planet’s atmosphere.

\section{Advantages of the Proposed Method and Future Directions}

\subsection{Advantages}

The proposed multi-object high-resolution observational strategy is expected to offer three key advantages:  

\noindent \textbf{1. Normalization of Blaze Function:} Our proposed simultaneous observation of a comparison star, similar in spectral type and magnitude to the program star, provides a direct empirical model of the combined instrumental response and stellar continuum. Dividing the program star’s spectrum by that of the comparison star could reduce the need for polynomial-fit–based continuum normalization. Also, this approach can remove the common-mode continuum shape and stellar contamination while preserving the planet’s absorption features.

\clearpage
\noindent \textbf{2. Enhanced Removal of Atmospheric Effects and Systematics:} Simultaneous observations of the program star and a comparison star with a high-resolution spectrograph allow for more effective correction of telluric and airmass variations, as well as instrumental systematics, particularly when combined with methods such as Gaussian Process Regression (\textsc{Gpr} )\footnote{\url{https://github.com/markfortune/luas}} or SYSREM\footnote{\url{https://pyastronomy.readthedocs.io/en/latest/pyaslDoc/aslDoc/aslExt_1Doc/sysrem.html}} \cite{Jiang,Birkby}.

\noindent \textbf{3. Combining \textsc{HRTS} \& \textsc{LRTS}:} This strategy enables the extraction of both \textsc{Hrts} and \textsc{Lrts} from the same dataset (see Figure~\ref{fig:hrts}). Combining ground-based \textsc{Hrts} with \textsc{Lrts} improves the constraints on atmospheric composition by jointly incorporating the likelihood functions of both approaches in retrieval analyses (see Section~\ref{retrieval}).

\subsection{Future Work and Implementation Plans}

The demonstrated high-resolution spectrophotometric method can also be applied to single-object observations by constructing spectroscopic light curves for each echelle order, after correcting for telluric lines, airmass trends, and other instrumental systematics using \textsc{Gpr}. 
This approach allows for an evaluation of the relative advantages and limitations of using a comparison star compared to single-object, and we will compare our results with those reported in previous studies.

We expect that the proposed strategy will enhance the detection of small variations in exoplanetary atmospheres, which reveal interactions between the planet and its environment. The main challenge, however, lies in variable fiber losses, which are time-uncorrelated. These can be mitigated using wavelet denoising techniques \cite{Chakrabarty_2019} from the observational data. 

In addition, this method provides a framework for incorporating fiber positioner-fed spectrographs into transmission spectroscopy studies. The multi-object data further contribute to understanding fiber-input injection variability and its implications for high-precision radial velocity measurements \cite{Manjunath}. 

\noindent \textbf{Retrieval:}  
In this study, we presented the spectral division method in the high-resolution framework. Therefore, we will perform a combined retrieval, where the $\chi^2$ function is used as the likelihood for both \textsc{Hrts}  and \textsc{Lrts} (obtained using the spectrophotometric and chromatic Rossiter-McLaughlin effect methods). Explicitly, the likelihood function is defined as  

\begin{equation}
\ln L_{\text{\textsc{Lrts} }}  = -\frac{1}{2} \chi^2 = -\frac{1}{2} \sum_{i} \frac{(y_i - y_{\text{model},i})^2}{\sigma_i^2}
\label{chi}
\end{equation}  

As a next step, we plan to apply a cross-correlation technique \cite{Snellen2010_HD209458b_winds,Brogi_2012} to the \textsc{Hrts}  data, followed by a joint retrieval of both \textsc{Hrts}  and \textsc{Lrts}. The two likelihood functions will then be combined\footnote{\url{https://github.com/esedagha/atmospheric_interpretations}} within a Markov Chain Monte Carlo (MCMC) framework \cite{Brogi_2012} to place quantitative constraints on atmospheric parameters.  

\begin{equation}
\ln L_{\text{\textsc{Hrts} }}  = -\frac{N}{2} \ln \left( \frac{1}{N} \sum_{n} \frac{g^2(n - s)}{\sigma_n^2} + \frac{\alpha^2}{N} \sum_{n} \frac{f^2(n)}{\sigma_n^2} - \frac{2\alpha}{N} \sum_{n} \frac{f(n)g(n - s)}{\sigma_n^2} \right)
\end{equation}

\begin{equation}
\ln L_{\text{tot}} = \ln L_{\text{\textsc{Lrts} }} + \ln L_{\text{\textsc{Hrts} }}
\label{eq3}
\end{equation}

where $\ln L_{\text{tot}}$ is the combined log-likelihood function, $\ln L_{\text{\textsc{Lrts} }}$ is the log-likelihood for \textsc{Lrts} , expressed in terms of the $\chi^2$ statistic, and $\ln L_{\text{\textsc{Hrts} }}$ is the log-likelihood for \textsc{Hrts} , represented through the cross-correlation formalism. Here, $N$ is the total number of data points, $y_i$ are the observed values with uncertainties $\sigma_i$, $y_{\text{model},i}$ are the model predictions, $f(n)$ and $g(n-s)$ represent the model and observed high-resolution spectra respectively, and $\alpha$ is the scaling factor between them.
\label{retrieval}

\section{Conclusion}

This study introduces and demonstrates the feasibility of Mo-HRTS as a promising approach for exoplanet atmosphere characterization. By simultaneously observing a target exoplanet host star and a comparison star with a high-resolution spectrograph, the method can mitigate systematics potentially more effectively than conventional single-object \textsc{Hrts}  techniques. Using simulations with the \textsc{Uves/Flames} spectrograph on the \textsc{VLT} for the well-studied ultra-hot Jupiter \textsc{wasp}-121 b, we showed that it is possible to construct both broadband and narrow-band transmission spectra from the same dataset, providing an alternative to the chromatic Rossiter–McLaughlin effect.  

Furthermore, we discussed how this dual-framework approach enables the combined retrieval of planetary atmospheric parameters, reduces normalization degeneracies, and improves constraints on atmospheric parameters. This capability can open a pathway to accurate and efficient exoplanet atmosphere studies, enhancing the capabilities of next-generation ground-based instruments for exoplanet characterization.
\acknowledgments 

Manjunath acknowledges the Very Large Telescope Distributed Peer Reviewers and the Gemini Time Allocation Committee for their insightful suggestions, which have greatly improved this work by encouraging a broader and more comprehensive perspective. We also sincerely thank the Indian Institute of Astrophysics and the Department of Science and Technology for their continuous support throughout this project. 

\normalsize
\setlength{\baselineskip}{0.8\baselineskip}
\bibliography{report} 
\bibliographystyle{spiebib} 

\end{document}